\documentclass[]{agujournal}
\draftfalse

\journalname{Geophysical Research Letters}
\usepackage{hyperref}
\hypersetup{
    colorlinks=true,
    linkcolor=blue,
    filecolor=magenta,      
    urlcolor=cyan,
}
\begin{document}

\title{An Updated Solar Cycle 25 Prediction with AFT: The Modern Minimum}

\authors{Lisa A. Upton\affil{1}, David H. Hathaway\affil{2}}

\affiliation{1}{High Altitude Observatory, National Center for Atmospheric Research, 3080 Center Green Dr., Boulder, CO 80301, USA}
\affiliation{2}{Solar Observatoires Group, Stanford University, HEPL-4085, Stanford, CA 94305-408, USA}

\correspondingauthor{Lisa Upton}{upton.lisa.a@gmail.com}

\begin{keypoints}
\item Cycle 25 will be slightly weaker than Cycle 24, making it the weakest cycle on record in the last hundred years.
\item Weak cycles are preceded by long extended minima -- we may not reach the Cycle 24/25 minimum until 2021.
\item We are currently (beginning with Cycle 24) in the midst of the modern Gleissberg cycle minimum.
\item It is too early to determine if this will remain a short Gleissberg minimum (like the Dalton) or if the Sun will produce a longer grand minimum (like the Maunder).
\end{keypoints}

%% \begin{abstract} starts the second page 

\begin{abstract}
Over the last decade there has been mounting evidence that the strength of the Sun's polar magnetic fields during a solar cycle minimum is the best predictor of the amplitude of the next solar cycle. Surface flux transport models can be used to extend these predictions by evolving the Sun's surface magnetic field to obtain an earlier prediction for the strength of the polar fields, and thus the amplitude of the next cycle. In 2016, our Advective Flux Transport (AFT) model was used to do this, producing an early prediction for Solar Cycle 25. At that time, AFT predicted that Cycle 25 will be similar in strength to the Cycle 24, with an uncertainty of about 15\% . AFT also predicted that the polar fields in the southern hemisphere would weaken in late 2016 and into 2017 before recovering. That AFT prediction was based on the magnetic field configuration at the end of January 2016. We now have 2 more years of observations. We examine the accuracy of the 2016 AFT prediction and find that the new observations track well with AFT's predictions for the last two years. We  show that the southern relapse did in fact occur, though the timing was off by several months. We propose a possible cause for the southern relapse and discuss the reason for the offset in timing. Finally, we provide an updated AFT prediction for Solar Cycle 25 which includes solar observations through January of 2018.

\end{abstract}

\section{Introduction}

The appearance of solar activity (sunspots, flares, coronal mass ejections, etc) is cyclic with an average period of about 11 years. Large solar storms, which also vary with the solar activity cycle, produce space weather events that can have devastating impacts on our assets in space, as well as here on Earth (e.g., communications and power grids). Accurate solar cycle predictions are essential for planning of future and current space missions and for minimizing disruptions to the nation's infrastructure.  

While there are still several different solar cycle prediction techniques \citep{2008Pesnell,2015Hathaway}, one method is emerging as a definitive leader in the field: the amplitude of the Sun's polar magnetic fields at solar cycle minimum  \citep [e.g.,][]{2005Svalgaard_etal, 2013MunozJaramillo_etal}. Surface Flux Transport (SFT) models  \citep{2005Sheeley,2014Jiang_etalB}, which simulate the evolution of the Sun's magnetic field, provide a way of estimating the amplitude of the polar fields several year prior to solar minimum, thereby extending the range of solar cycle predictions.

The Advective Flux Transport (AFT) model is one such SFT model, designed specifically with the intent of being as realistic as possible without the use of free parameters \citep {2014UptonHathawayA, 2014UptonHathawayB, 2015UgarteUrra_etal}. The AFT model was recently used to make an ensemble of 32 predictions for the amplitude of Solar Cycle 25  \citep {2016HathawayUpton} (hereafter referred to as HU16). In this study, 3 model parameters - the convective motion details, active region tilt, and meridional flow profile - were varied in order to also determine the relative uncertainty produced. HU16 found that the polar fields near the end of Cycle 24 would be similar to or slightly smaller than the polar fields near the end of Cycle 23, suggesting Cycle 25 would be similar or somewhat weaker than Cycle 24. After four years of simulation, the variability across the ensemble produced an accumulated uncertainty of about 15 \%. Additionally, all realizations in the HU16 ensemble predicted a relapse in the southern polar field in late 2016 and into 2017. 

One of the biggest sources of uncertainty in making solar cycle predictions comes from the large scatter inherent in the systematic (Joy's Law) tilt of Active Regions (ARs)\citep{2014Jiang_etalA}. This tilt angle produces an axial dipole moment in newly emerged ARs, which continues to evolve during the lifetime of the AR. Over the course of the solar cycle, the axial dipole moments of the residual ARs  are transported to higher latitudes, where they accumulate, causing the reversal and build up of the polar fields. The net global axial dipole at the end of the cycle (i.e., solar cycle minimum), forms the seed that determines the amplitude of the next cycle.

 \citet{2014Cameron_etal, 2017Nagy_etal} showed that large, highly tilted 'rogue' active regions can have a huge impact on the Sun's axial dipole moment, particularly if they emerge close to the equator. We are now two years closer to solar minimum since our last prediction. At this late stage of the solar cycle, fewer ARs emerge, reducing the likelihood that a 'rogue' active region will emerge. Those that do emerge typically have much weaker flux \citep{2015MunozJaramillo}, emerge closer to the equator (Sp\"orer's Law), and have smaller tilt angles (Joy's Law). The net effect of all of these factors, barring the emergence of a large 'rogue' active region, means that the few ARs that are left to emerge will have very small axial dipole moments and little impact on the polar field strengths. Another effect is that the uncertainly caused by the variability in the tilt is significantly reduced. With the solar cycle minimum only 2-3 years away, this is an optimal time for an updated prediction.   

In this paper we begin by revisiting the previous Solar Cycle 25 prediction made with the AFT model. We discuss the accuracy of those predictions as compared to the observations that have since occurred. We then provide an updated prediction for Solar Cycle 25.

\section{Previous Prediction Fidelity}

 \begin{figure}[t]
 \centering
 \includegraphics[width=30pc]{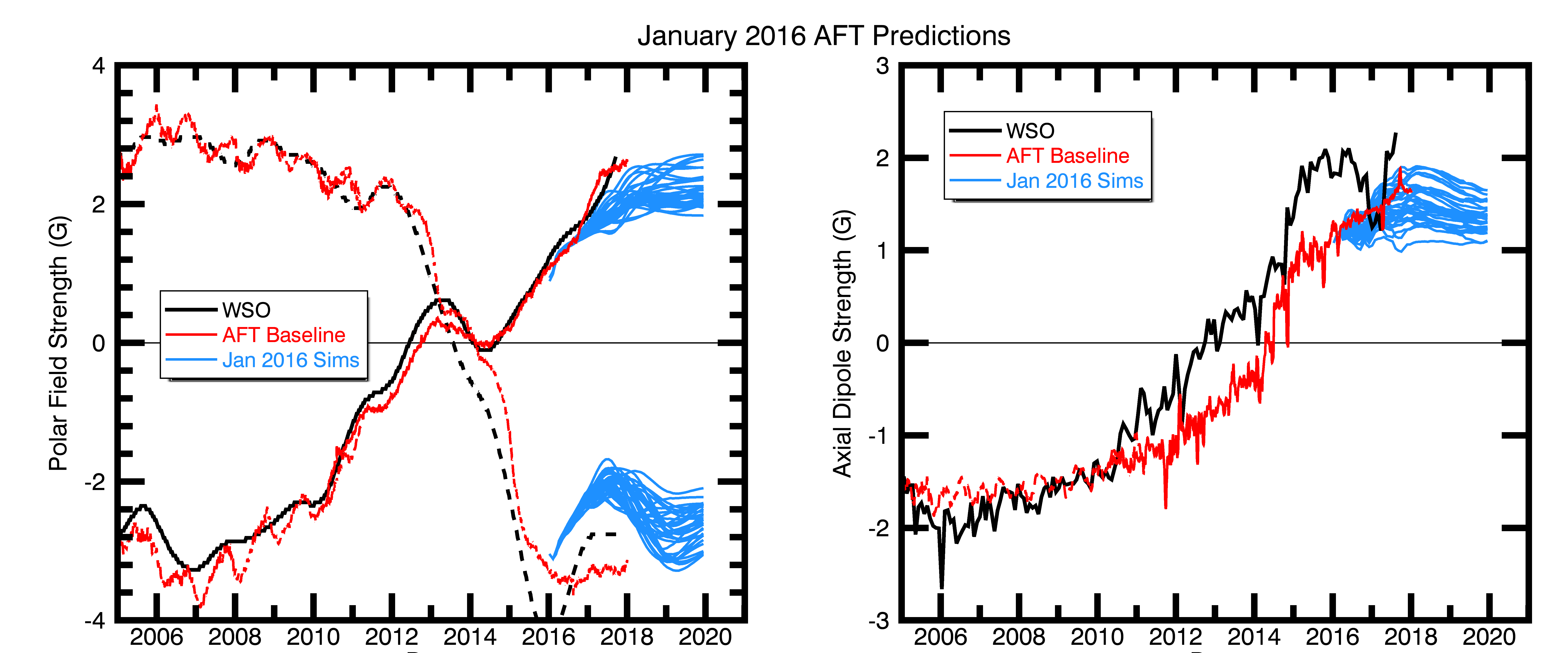}

 \caption{Validating the AFT 2016 Predictions. This figure shows the polar field predictions that were made in \citep {2016HathawayUpton} (in blue) along with the polar field observations (WSO in black and AFT Baseline in red) that have occurred since the prediction was made. The average polar fields poleward of 55\textdegree are shown on the left. The polar field strength as measured from the axial dipole moment is shown on the right.}
 \label{fig:Jan2016}
  \end{figure}

The prediction of HU16 was initiated in January 2016. We now have 2 years of observations of the Sun's polar fields to compare with and investigate the accuracy of those simulations. We begin with the predicted and observed axial dipole moment (Figure \ref{fig:Jan2016}, right panel) and find that the observations track right in the middle of the ensemble of predictions. Next, we compare the polar fields as measured above 55\textdegree  (Figure \ref{fig:Jan2016}, left panel). In the northern hemisphere, the observations track the predictions fairly well. There is a strong agreement for the first year, but the predictions are slightly weaker in the second year. However, when we compare the polar fields in the southern hemisphere, we find that the agreement is not as good. While the southern hemisphere relapse that was predicted in HU2016 did in fact occur, it appears to happen about nine months later than the prediction. \textbf{Why did the southern relapse occur later in the observations than in AFT predictions?} Before we can answer this question, we must first look at the reason that the southern relapse occurred in the first place.

 \begin{figure}[t]
 \centering
 \includegraphics[width=33pc]{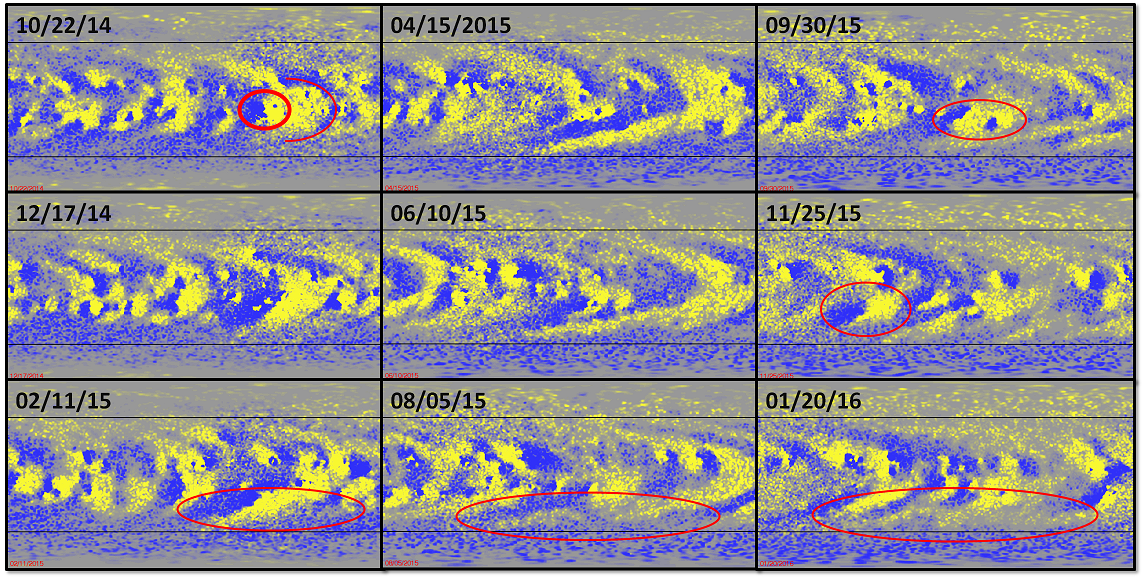}

 \caption{Sequence of AFT magnetic Maps. This sequence of AFT maps has been supersaturated to enhance the appearance of the weak magnetic field at the poles. The 55\textdegree latitude lines have been marked with thin black lines. AR 12192 has been circled in red on the top left panel. The subsequent evolution of AR 12192 can been seen in the first two rows. The red circled regions in the bottom panels show the formation of a positive polarity region right at the55\textdegree latitude. Two additional active regions, 12415 (top right panel) and 12422 (middle right panel), occurred later and also contributed to the formation of the positive polarity region at the 55\textdegree \space latitude line.
}
 \label{fig:AFTmaps}
  \end{figure}

The updated observations show that the southern polar field started off progressing normally, with the negative polarity growing. But then, in October of 2014, a new extraordinary Active Region emerged, AR 12192, which created a very large positive polarity stream which was transported to the South, as shown in Figure \ref{fig:AFTmaps}. AR 12192 had Hale's polarity and a small tilt angle, consistent with Joy's Law. It was the Largest Active Region in the last 24 years and it ranked 33rd largest of 32,908 active regions since 1874 \citep{2015Sun_etal}. From the sequence of maps in Figure \ref{fig:AFTmaps}, we see that both the leading and following polarities are sheared out by the differential rotation. Both polarities are transported to high latitudes, but this shearing effect pushed the leading positive polarity flux to higher latitudes than the negative following polarity. The polar-effectiveness (e.g, the amount of flux transported to the poles) of the leading polarity may have been enhanced because is was surrounded by a weak negative flux region. This  minimized cancellation at the 'Bow'-side of the AR, while the following polarity flux was squeezed and canceled by the positive polarity flux that surrounded it. This culminated in a weak band of positive polarity flux just above the 55\textdegree line in the South, as seen in the bottom middle panel of Figure \ref{fig:AFTmaps}. But AR 12192 only laid the foundation for the relapse. The positive band it created was aided by subsequent Active Regions (most notably NOAA 12415 and NOAA 12422), which helped to enhance the positive polarity band that formed in the South. As this positive polarity band progressed poleward, it degraded the negative southern pole, causing the subsequent relapse.

Strong shear in the differential rotation at mid latitudes stretches the magnetic flux in the East-West direction. When both polarities are transported to high latitudes, this tends to produce alternating bands of flux which form long polarity inversion lines stretching East-West. Throughout 2016, the neutral line for the positive band was right at 55\textdegree latitude. This latitude, coincidently, was the cutoff used to measure the hemispheric polar fields (Figure \ref{fig:Jan2016}, left panel). Small differences in the SFT processes (e.g., meridional flow and convection pattern) can shift significant amounts of flux above or below this arbitrary line. This can cause differences in the polar field measurements above that latitude. This difference in the timing of the flux crossing 55\textdegree  translates into a difference in the timing of the relapse. The HU2016 simulations had slightly more of the positive flux cross the 55\textdegree line, causing the relapse to occur sooner in those simulations. This serves as a reminder that while polar field measurements above a given latitude are useful for identifying hemispheric asymmetries, they can be somewhat subjective and lead to offsets in prediction timing \citep {2014UptonHathawayA}. Despite this offset in the timing, we are reassured by the fact that the axial dipole predictions (Figure \ref{fig:Jan2016}, right panel) are remarkably well matched - falling within the middle of the prediction ensemble. This provides confidence in the ability of AFT to accurately predict the evolution of the polar fields at least two years in advance during the early part of the declining phase of the solar cycle.

\section{Updated Cycle 25 Prediction}

\begin{figure}[t] 
 \centering
 \includegraphics[width=30pc]{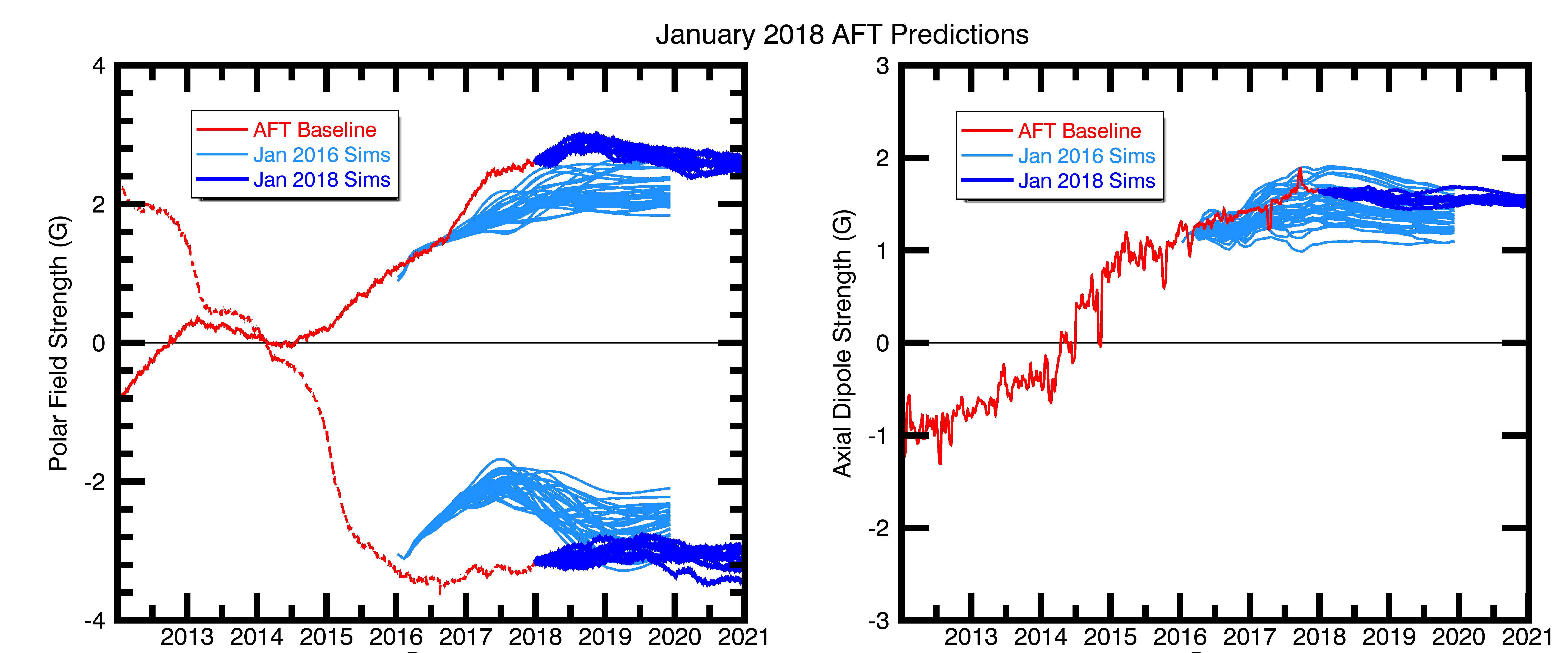}

 \caption{AFT 2018 Predictions. This figure shows the polar field observations (red) along with the AFT predictions (2016 in the lighter blue and 2018 in the darker blue). The polar fields strengths as measured from 55\textdegree and above are shown on the left. The polar field strength as measured from the axial dipole moment is shown on the right.}
 \label{fig:AFT2018}
  \end{figure} 
  
We now have two additional years of observations, since the predictions of HU2016. Here, instead of a start date Jan 2016, we'll start the new prediction at Jan 2018. At this time, the northern polar is stronger, and the southern polar field is weaker than they were in Jan of 2016. As it is later in the cycle, we expect fewer active regions to emerge. The active regions that do emerge will be smaller, will be at lower latitudes and will tend to have a small tilt angle. All of these characteristics work together to reduce the axial dipole moment of each active region, thereby reducing its polar effectiveness. At this late stage of the cycle, the active regions that will emerge will have little to no effect on the polar fields that will ultimately produce Solar Cycle 25, significantly minimizing the uncertainty in our prediction for the next cycle.  

Here we ran 10 simulations using the active regions from solar cycle 14, varying both Joy's tilt and the convective pattern (see HU2016 for the details). The results of all of these simulations are shown in Figure \ref{fig:AFT2018}. The average of all 10 realizations gives an axial dipole strength at the start of 2020 of +1.56 $\pm 0.05$ G. WSO gave an axial dipole strength of -1.61 G at the start of Cycle 24, +3.21 G at the start of Cycle 23, and -4.40 G at the start of Cycle 22. \textbf{This suggests that Cycle 25 will be a another small cycle, with an amplitude slightly smaller than ($\sim$ 95-97\%) the size of Cycle 24. This would make Solar Cycle 25 the smallest cycle in the last 100 years.} This indicates that the weak cycle 24 is not an isolated weak cycle, but rather the onset of the modern Gleissburg minimum \citep {1939Gleissberg}, which will include Cycle 25 \textemdash \, at present this is akin to the last Gleissburg minimum (SC12, SC13, \& SC14) which occurred in the late 1800s and early 1900s. Unfortunately, we will need to wait another 10-15 years before we will know if the Sun will go into a deeper minimum state (e.g. the Dalton or Maunder minima, or somewhere in between) or if it will recover as it did following the last Gleissberg minimum.

Weak cycles are preceded by long extended minima \citep{2015Hathaway} and we expect a similar deep, extended minimum for the Cycle 24/25 minimum in 2020. Based, on the latest prediction, \textbf{we expect that minimum will be closer to the end of 2020 or beginning of 2021.} Long extended minima such as this are punctuated by a large number of spotless days (e.g., SC12-SC15 and SC24).  Similarly, \textbf{we expect that the Cycle 24/25 minimum will include extended periods of spotless days throughout 2020 and into 2021.} Fortunately, the strength of the axial dipole doesn't change much during 2020: +1.56 $\pm 0.05$ G for the start of 2020 and +1.54 $\pm 0.04$ G for start of 2021. Therefore, this extended minimum should have little impact on the prediction for Cycle 25.

\section{Conclusions}

We have investigated the accuracy of the predictions made by AFT in 2016 (HU2016). We found that those predictions are largely in line with the observations that have occurred since that prediction was made.The biggest discrepancy was found to be the timing of a relapse in the strength of the southern polar field - while the amplitude was correct, the relapse actually occurred about 9 months later. We identified a few active regions that produced leading polarity streams that caused this relapse, with the most significant of these ARs, being NOAA 12192. We found that the offset in the timing of the relapse was due primarily to formation of the polarity inversion line right at the 55\textdegree latitude cutoff. Slight differences in the surface flux transport can significantly change the amount of flux above or below this line, resulting in offsets in the timing of the evolution of the hemispheric polar fields. Despite this offset, the evolution of the axial dipole for the last 2 years was accurately predicted in HU2016. 

We provided an updated prediction for solar Cycle 25, which incorporated the observations up to Jan 2018. The new prediction gave an axial dipole of +1.56 $\pm 0.05$ G for the start of 2020 and +1.54 $\pm 0.04$ G for start of 2021. This indicated that Cycle 25 will be on the order of 95\% of Cycle 24. Of the predictions that are using the axial dipole as a predictor, AFT is on the lower end of the spectrum. \citep {2017Jiang_Cao} expects the axial dipole at 2020 to be 1.76 $\pm$ 0.68 G, or comparable to Cycle 24. \citep {2017Wang} also expects Cycle 25 to be comparable to Cycle 24. \citep {2016Cameron_etal} predicts that Cycle 25 will be slightly higher than Cycle 24, but acknowledges that the reliability of this prediction is limited by the intrinsic uncertainty. Given the consensus of these predictions with our own results, we are confident that Cycle 25 will indeed be another weak cycle. 

We note that our new prediction ( +1.56 $\pm 0.05$ G) falls within the uncertainty given in our HU2016 prediction (+1.36 $\pm 0.20$ G). While this demonstrates that AFT can accurately predict the evolution of the axial dipole, within the uncertainty, 4 years in advance of the minimum, the addition of two more years of observations significantly adds to the precision of the AFT solar cycle predictions. At this late stage of the cycle, the uncertainty in AFT's ability to predict the polar fields is very small. We acknowledge that there is additional uncertainty associated with using the axial dipole as a predictor of the amplitude of the next cycle. Compounding this is the fact that, while this trend appears to be linear for cycles stronger than Cycle 24, we do not yet have data to show that this relationship holds for cycles that are weaker than Cycle 24 (see Figure 1 HU2016, which shows Cycle 24 is the smallest cycle used to determine this relationship). Though we do make this assumption in our prediction for the strength of Cycle 25, Cycle 25 will be a test of this assumption. As the saying goes, \textit{only time will tell}, but we await it with open arms.  

\acknowledgments

The data presented in this article are freely and publicly available at the following web address: \url{http://solarcyclescience.com/Predictions/2018GRLData.zip}. 

L.A.U. was supported by the National Science Foundation Atmospheric and Geospace Sciences Postdoctoral Research Fellowship Program (Award Number:1624438) and is hosted by the High Altitude Observatory at National Center for Atmospheric Research (NCAR) . NCAR is sponsored by the National Science Foundation. We would like to thank Robert Cameron for insightful discussions about the Southern Relapse. Finally, we would like to thank the anonymous referees for their careful reading of our manuscript and their valuable comments and suggestions.

\bibliography{main.bib}

%
%%%%%%%%%%%%%%%%%%%%%%%%%%%%%%%%%%%%%%%%%%%%%%%
% Or, to use BibTeX:
%
% Follow these steps
%
% 1. Type in \bibliography{<name of your .bib file>} 
%    Run LaTeX on your LaTeX file.
%
% 2. Run BiBTeX on your LaTeX file.
%
% 3. Open the new .bbl file containing the reference list and
%   copy all the contents into your LaTeX file here.
%
% 4. Run LaTeX on your new file which will produce the citations.
%
% AGU does not want a .bib or a .bbl file. Please copy in the contents of your .bbl file here.

%% After you run BibTeX, Copy in the contents of the .bbl file here:

%%%%%%%%%%%%%%%%%%%%%%%%%%%%%%%%%%%%%%%%%%%%%%%%%%%%%%%%%%%%%%%%%%%%%
% Track Changes:
% To add words, \added{<word added>}
% To delete words, \deleted{<word deleted>}
% To replace words, \replace{<word to be replaced>}{<replacement word>}
% To explain why change was made: \explain{<explanation>} This will put
% a comment into the right margin.

%%%%%%%%%%%%%%%%%%%%%%%%%%%%%%%%%%%%%%%%%%%%%%%%%%%%%%%%%%%%%%%%%%%%%
% At the end of the document, use \listofchanges, which will list the
% changes and the page and line number where the change was made.

% When final version, \listofchanges will not produce anything,
% \added{<word or words>} word will be printed, \deleted{<word or words} will take away the word,
% \replaced{<delete this word>}{<replace with this word>} will print only the replacement word.
%  In the final version, \explain will not print anything.
%%%%%%%%%%%%%%%%%%%%%%%%%%%%%%%%%%%%%%%%%%%%%%%%%%%%%%%%%%%%%%%%%%%%%

%%%
\listofchanges
%%%

\end{document}